\documentclass[10pt,twocolumn]{article}      

\usepackage{soul}

\usepackage{times}
\usepackage[top=18mm,left=22mm,right=22mm,bottom=25mm,columnsep=5mm]{geometry}

\usepackage{cite}
\usepackage[table,usenames,dvipsnames]{xcolor}
\usepackage{hyperref}
\usepackage{rotating}
\usepackage{color, colortbl}
\usepackage{balance}
\usepackage{multirow}
\definecolor{Gray}{gray}{0.9}
\usepackage{graphicx} 
\graphicspath{ {./Figs/} }
\usepackage{wrapfig}
\usepackage{comment}
\usepackage{subcaption}
\usepackage{mdframed}
\usepackage{tikz}
\usepackage{dirtytalk}
\newcommand*\circled[1]{\tikz[baseline=(char.base)]{
            \node[shape=circle,draw,inner sep=2pt] (char) {#1};}}

\title{
Making Sense of Failure Logs in an Industrial DevOps Environment
}

\usepackage{authblk}
\author[1,2]{M Abbas}
\author[2]{A Hamayouni}
\author[1]{M H Moghadam}
\author[1]{M Saadatmand}
\author[3]{P E Strandberg}
\affil[1]{\normalsize{RISE Research Institutes of Sweden, V\"aster\aa s, Sweden}}
\affil[2]{\normalsize{M\"alardalens University, V\"aster\aa s, Sweden}}
\affil[3]{\normalsize{Westermo Network Technologies AB, V\"aster\aa s, Sweden}}

\date{\normalsize{Preprint, accepted to ITNG 2023}}

\begin{document}

\maketitle

\section*{Abstract}
Processing and reviewing nightly test execution failure logs for large industrial systems is a tedious activity.
Furthermore, multiple failures might share one root/common cause during test execution sessions, and the review might therefore require redundant efforts.
This paper presents the LogGrouper approach for automated grouping of failure logs to aid root/common cause analysis and for enabling the processing of each log group as a batch.
LogGrouper uses state-of-art natural language processing and clustering approaches to achieve meaningful log grouping.
The approach is evaluated in an industrial setting in both a qualitative and quantitative manner.
Results show that LogGrouper produces good quality groupings in terms of our two evaluation metrics (Silhouette Coefficient and Calinski-Harabasz Index) for clustering quality.
The qualitative evaluation shows that experts perceive the groups as useful, and the groups are seen as an initial pointer for root cause analysis and failure assignment.

\phantom{space}

\noindent
KEYWORDS:
failure clustering, nightly testing, failure embedding, root cause analysis, DevOps, test logs, log analysis, software testing, word cloud, natural language processing

\section{Introduction}\label{sec:intro}
Incremental development of large industrial software systems necessitates continuous integration and testing.
The repeated execution of regression test suites on different versions and variants of these systems produces a huge number of test execution logs that must be reviewed by engineers~\cite{golagha2019failure} \cite{eljasik2020leveraging}. 
As an example, our industrial partner Westermo Network Technologies AB (Westermo) develops the Westermo Operating System (WeOS)\footnote{WeOS, available online, \url{https://www.westermo.com/solutions/weos}} for different industrial switches and routers. 
WeOS contains the Linux kernel, other free software, as well as proprietary code.
Each new feature, such as an extension of a communication protocol, is developed in a separate feature branch that receives significant quality assurance prior to being merged into the main branch. 
In addition to manual testing and debugging, WeOS is tested nightly in several test systems built up of devices forming different network topologies.
This produces a huge amount of test execution logs---often duplicates or very similar to other ones in the batch---that has to be reviewed.
Typically, engineers review critical warnings and failures caused during the test execution.
This creates significant overhead for developers and test engineers because most of the failure logs are duplicates or share a common cause.
In most cases, the engineers might miss diverse failures at the end of the long list, as the engineer might be under the impression that most of the logs are the same.

Failures logs contain textual information, and based on the similarity, they can be grouped using clustering approaches. 
The application of clustering and other Natural Language Processing (NLP) approaches could be leveraged to assist engineers in nightly log reviews.
Failure clustering could aid the log processing and root cause analysis in the DevOps environment and even assist test managers in failure assignment.
Recent studies have attempted to find a way to improve feature extraction methods and boost the logs representation accuracy to generate acceptable clusters/groups.
Several studies in this area have used NLP-based approaches such as the term frequency inverse document frequency (tfidf) and support vector machine models~\cite{eljasik2020leveraging}, key extraction~\cite{fu2009execution}, word embedding~\cite{bertero2017experience}, and Latent Semantic Analysis (LSA) techniques~\cite{digiuseppe2012concept}. 
Literature reveals that log clustering directly impacts the speed and performance of failures logs analysis (e.g.,~\cite{eljasik2020leveraging}). 
However, most studies above investigated failure clustering from a quantitative standpoint and rarely in an industrial setting.

\textit{Contributions.}  We report an approach called LogGrouper that leverages state-of-the-art NLP techniques in combination with various clustering algorithms to group failure logs.
LogGrouper aims to assist engineers in reviewing nightly failure logs by providing a holistic view of the types (categories) of the occurring failures.
Furthermore, we evaluate the different clustering approaches---supported by LogGrouper---for failure log clustering based on clustering quality metrics in an industrial setting.
Finally, we also evaluate the LogGrouper approach using qualitative data obtained through a focus group study conducted with eleven experts at Westermo.
Results show that the LogGrouper approach produced meaningful failure groups with Density-based spatial clustering of applications with noise (DBSCAN) clustering algorithm performing the best in terms of clustering quality.
Qualitative results show that LogGrouper failure groups are useful and could act as an initial pointer for root cause analysis and failure assignment.

\textit{Structure.} The rest of the paper is organized as follows. 
Section~\ref{sec:bg} introduces the background and related work briefly. 
Section~\ref{sec:approach} presents the approach. 
Section~\ref{sec:method} presents the evaluation method, Section~\ref{sec:results} presents and discusses the obtained qualitative and quantitative results.
Section~\ref{sec:validity} presents the potential threats to validity and limitations.
Finally, Section~\ref{sec:conc} concludes the paper with future directions.

\section{Related work}\label{sec:bg}
Various studies have focused on failure clustering and log analysis. The following sections discuss several recent related works in this area.

The use of machine learning for log embedding and analysis is a common approach to aid log grouping. In this regard, Sharp \textit{et al.}~\cite{sharp2016semi} used the Bag of Words Model and Word2vec to propose a data-driven tags method for logs.
They also clean and prepare data before assigning tags to them. The proposed method also recognizes recurrent long-term logs in addition to clustering data into separate groups.
Xiao \textit{et al.}~\cite{xiao2020lpv} used word2vec to represent textual logs and devised a log mining method called Log parser based on Vectorization (LP). 
This study is primarily concerned with offline and online log parsing. 
The similarity between two logs is calculated using the extracted log vectors in offline log parsing. While in online parsing, the approach uses average vectors as log templates to detect similarities between incoming log messages.

Eljasik \textit{et al.} proposed the Log Analysis Machine Learner (LAMaLearner) system to cluster unstructured text log files that do not have standard information~\cite{eljasik2020leveraging}. 
They use various Artificial Intelligence (AI) approaches to find the acceptable logs features representation, such as combining the Support Vector Machine (SVM) model with tfidf vectors. 
Aussel \textit{et al.}~\cite{aussel2018improving} and Bertero \textit{et al.}~\cite{bertero2017experience} also rely on NLP to extract features from unstructured log files in order to provide adequate input for log methods such as analysis, mining, anomaly detection, and classification. 
These studies analyze the influence of NLP-based feature extraction strategies on log analysis using a variety of datasets and logs.

Itkin \textit{et al.}~\cite{itkin2019user} presented a practical presentation of identifying and categorizing, and cleaning settlement system failure logs. 
This study uses several data mining approaches, such as hierarchical agglomerative clustering with brute-force comparison and K-means with factorized tfidf word embedding, to form a log analysis framework.
Fu \textit{et al.}~\cite{fu2009execution} propose using key extraction to cluster and detect anomalies in pure text of logs. The variable parts of log messages are eliminated in this study via predefined rules, which means that the proposed approaches require some rudimentary information about the input dataset.
Lin \textit{et al.}~\cite{lin2016log} proposed a log partitioning strategy for clustering and identifying log occurrence patterns, called LogCluster.
Their results show that with the use of LogCluster, the log inspection process has become quicker, and LogCluster could discover previously undiscovered logs in the process. 

Literature also reveals a handful of secondary studies on log clustering and analysis. He \textit{et al.}~\cite{he2016evaluation} reviewed recent work on log parsers and compared and analyzed their performance when confronted with a vast amount of logs as input.
They concluded that present log parsers are unsuitable for huge datasets, and the majority of them lack the requisite execution time efficiency.
Korzeniowski\textit{et al.}~\cite{korzeniowski2022landscape} conducted a systematic mapping study on the log analysis research.
They presented their findings in the form of a landscape of automated log analysis, defining each discipline and highlighting recent research trends.

Existing studies mainly investigated failure clustering from a quantitative standpoint and rarely in an industrial setting. In this paper, we explore the problem from a qualitative and quantitative standpoint in an industrial setting.

\begin{figure*}[!th]
      \centering
      \includegraphics[width=0.75\linewidth]{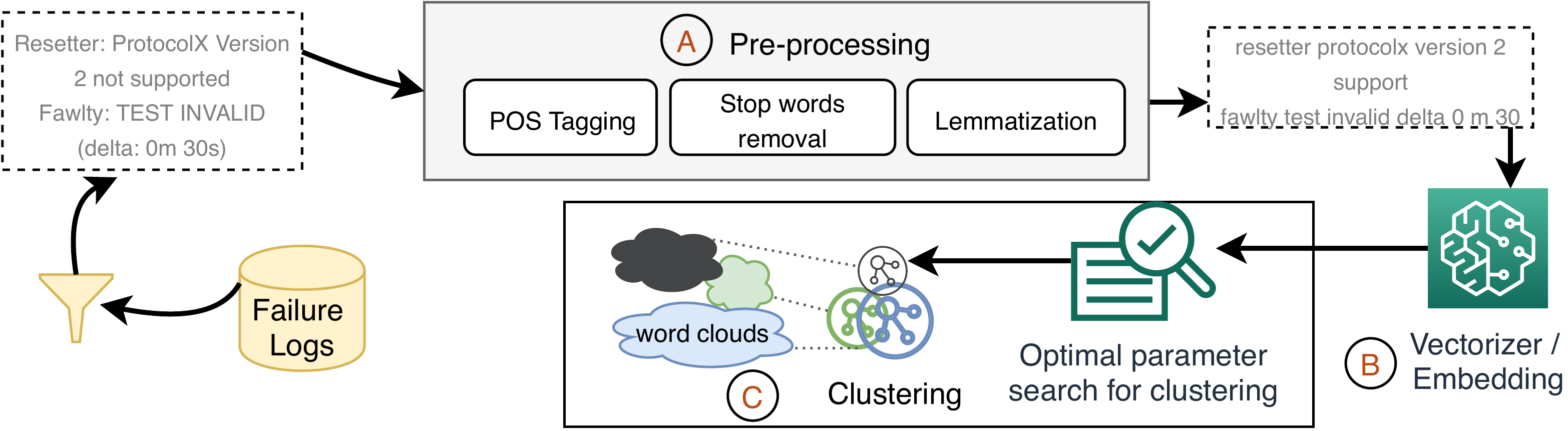}
      \caption{LogGrouper approach for grouping failure logs of regression tests}
      \label{fig:approach}
 \end{figure*}
\section{LogGrouper --- Approach}\label{sec:approach}
In this section, we present the LogGrouper approach for failure log grouping in a DevOps environment. 
As shown in Figure~\ref{fig:approach}, LogGrouper has three distinct steps that are packaged as a Representational State Transfer (REST) Application Programming Interface (API).
As an initial step, LogGrouper filters the relevant parts of log files (errors and critical warnings) from the log repository and uses it as input. 
The failure logs are then passed through a pre-processing pipeline---step \circled{A} in Figure~\ref{fig:approach}---, where the content of the failure logs is cleaned for further steps.
The cleaned text of the failure logs is then converted into numerical feature vectors---in step \circled{B} of Figure~\ref{fig:approach}---to be used for clustering.
Before clustering the failure logs, we compute the optimal parameters---such as the number of clusters through \textit{elbow method} for k-means clustering---based on the feature vectors inferred from the failure logs~\cite{dangeti2017statistics}.
LogGrouper then clusters the failure logs---both with and without pre-processing applied---in multiple clusters (groups) in step \circled{C}, as shown in Figure~\ref{fig:approach}.
As a next step, LogGrouper evaluates a variety of clustering approaches for each vectorization approach and outputs the results of the best clustering algorithm in terms of the quality of the clusters.
However, the clusters alone might not be easily interpretable by end-users. Therefore, LogGrouper uses a key-phrase extraction algorithm to extract key phrases from each group of failure logs that are visualized in the form of word clouds.
Below, we present each step of LogGrouper in detail.

\textit{A. Pre-Processing:}
The noisy nature of log messages may complicate the downstream NLP tasks. 
Log files contain multiple attributes that are not useful for log clustering.
As an initial step toward pre-processing, we convert the text of the log messages into lowercase and remove timestamps, symbols, and indentations.
To avoid the interpretation of the same terms in different language forms, we lemmatize each token of the log files.
This may help in meaningful similarity computation in later tasks.
We have shown an example part of the log message before and after pre-processing in the text boxes with dotted borders in Figure~\ref{fig:approach}.

\textit{B. Feature vectors extraction:}
Feature vector extraction from textual data is typically done by mapping the text into vector representation in the vector space model.
This is achieved by first learning a representation from the data as a language model and then inferring the feature vectors for the log messages. 
Therefore, such approaches are also known as language models, representation learning, vectorization, or word embedding approaches.
However, with the rise of transfer learning, large pre-trained language models have been making their way into software engineering tasks.
In this regard, LogGrouper uses a variety of different vectorization approaches together with pre-trained language models to vectorize the log messages for grouping/clustering log files.
Currently, LogGrouper supports the information retrieval (IR) model tfidf, the machine learning-based FastText~\cite{25FastText2016}, and the deep learning-based Bidirectional Encoder Representations from Transformer (BERT)~\cite{devlin2018bert}.

\textit{Feature vector's dimensionality reduction for IR:} The IR approach used by LogGrouper is based on extracting term-document matrix with tfidf scores as values for the features.
The vector extracted from the matrix results in sparse vectors with a large number of zeros.
Therefore, LogGrouper normalizes the vectors and uses the popular Principal Component Analysis (PCA) approach for dimensionality reduction~\cite{abdi2010principal}.
This results in a reduced vector of a reasonable size of dimensions.

\textit{C. Clustering \& its visualization:}
We consider four widely used clustering algorithms---for grouping failure logs---from three diverse classes of algorithms, as follows.
\textit{Agglomerative} is a bottom-up hierarchical clustering technique that begins by treating each item as a separate cluster and then repeats finding the two most similar clusters and merging them until a pre-defined number of clusters is reached. K-\textit{means} is a partitioning (centroid-based) clustering
algorithm that starts with a (K) set of randomly chosen centroids that serve as the cluster’s starting points. Then, it iteratively computes the sum of the squared distance between the data points and the cluster’s centroids, assigns each item to the closest centroid, and computes the new centroid for each cluster until centroids converge. Density-Based Spatial Clustering of Applications with Noise (\textit{DBSCAN}) gathers items close together into a cluster around a core point based on the distance between them. \textit{Spectral} is a clustering
technique that reduces the dimensions of the input vectors using
the Laplacian matrix before the actual clustering and then
uses other algorithms for clustering.

LogGrouper uses the aforementioned algorithms for clustering and also employs different strategies to find the best parameters for each clustering algorithm. 
In K-Means, agglomerative clustering, and Spectral, the Elbow method~\cite{dangeti2017statistics} is used for plotting the sum of squared distances as a function of \textit{k} and selecting the curve's elbow as the best \textit{k} in terms of Calinski Harabasz and Silhouette metrics (see Section~\ref{sec:method}).
For DBSCAN, Epsilon---(eps), the maximum distance between two items for treating one of them as a neighbor of the other---and minimum samples---the minimum number of items required to construct a cluster---are the two important parameters to be computed.
LogGrouper sets the minimum sample based on the recommendations in the literature~\cite{sander1998density} and uses the greedy search for computing the epsilon. 
After clustering, the approach outputs the results of the best clustering algorithm based on Calinski Harabasz and Silhouette metrics.
However, results from all other clustering pipelines could also be retrieved. Key phrases from all the clusters are separately extracted using the Rapid Application Keyword Extraction (RAKE) algorithm~\cite{rose2010automatic}. The key phrases are then visualized as word clouds.

\section{Evaluation}\label{sec:method}
\textit{A. Context and Research Questions:}
This work can be classified as an \textit{exploratory} case study that focuses on improving the log processing at Westermo.
The overall objective is to aid the root/common cause analysis of the nightly test execution logs and to aid the bulk processing of failure logs.
LogGrouper is a step towards achieving the main objective.
To evaluate the LogGrouper approach, we address the following research questions (RQs).\\
\textit{\textbf{RQ1}:} What clustering approaches result in high-quality grouping of failure log messages, w.r.t the evaluation metrics?\\
Since the quality of grouping is very much dependent on the clustering approaches used, we aim to identify the best clustering approach both with and without preprocessing with respect to evaluation metrics in the context of our dataset.
Note that the evaluation of the quality of vectorization approaches is out of this study's scope and will be considered in the future.\\
\textit{\textbf{RQ2}:} What are the perceptions of the engineers regarding the LogGrouper results?\\
In this RQ, we aim to gather qualitative insights on the perceptions of engineers regarding the results of LogGrouper, the perceived correctness of the groupings, and the challenges associated with the approach and its adoption.

\textit{B. Data Collection:}
Westermo provided test results data from two months of nightly testing to design and evaluate the approach.
The data was in the form of a test results database, log files from their nightly testing, as well as a tool for exploring and querying the test results~\cite{strandberg2022software}.
To evaluate the approach in the studied industrial setting, we use log messages that are tagged as erroneous and critical from three time windows that engineers might typically use in daily work for log review, as follows.
The first time window of one nightly session on one or more code branch(es), the second time window for one or many code branch(es), and finally, the third time window considers all the available log files.
This mimics a typical nightly test execution log review process.
Besides, we also use the generated grouping results, key phrases, and word clouds for qualitative evaluation of the approach.
\begin{figure}[!t]
      \centering
      \includegraphics[width=0.85\linewidth]{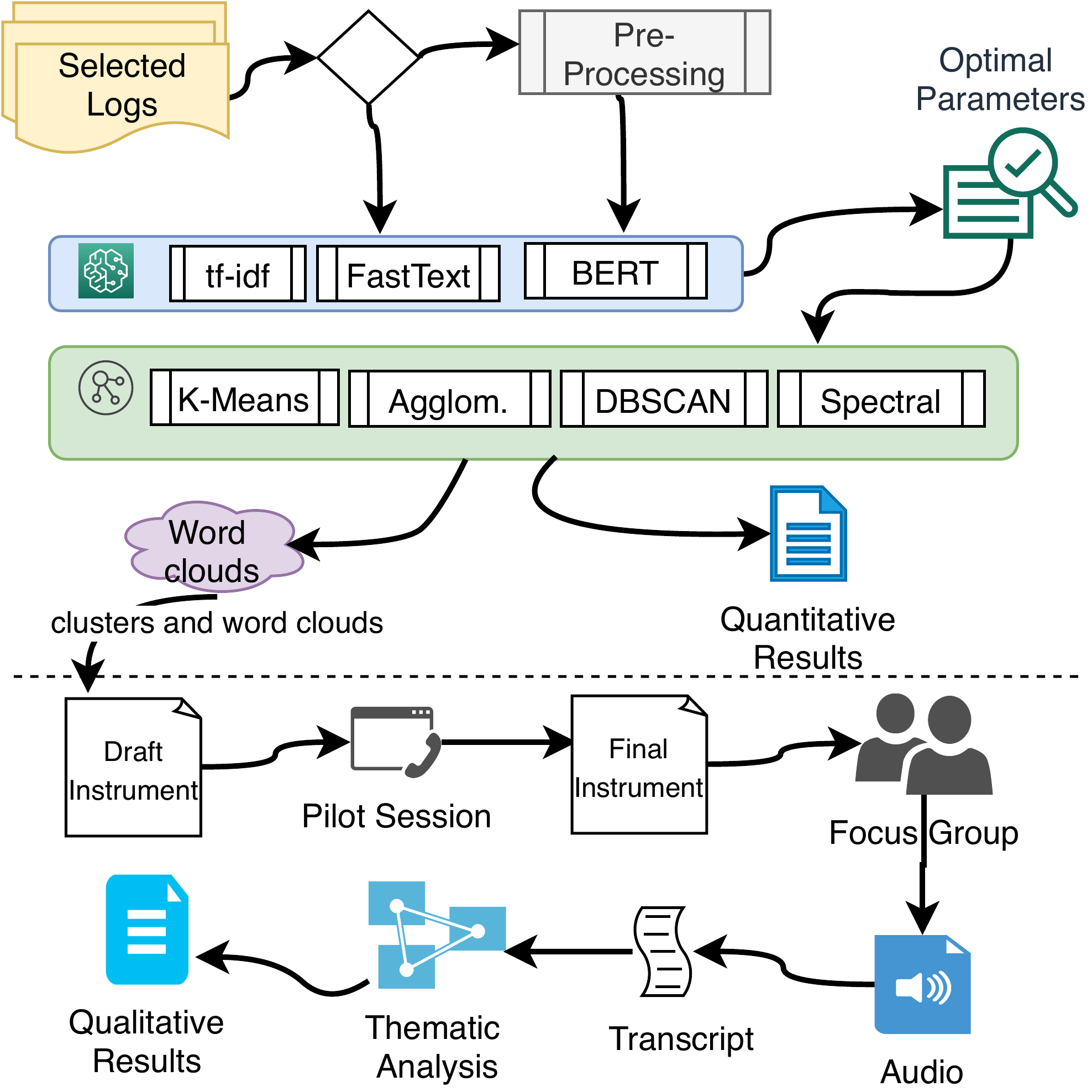}
      \caption{Evaluation procedure}
      \label{fig:procedure}
 \end{figure}

\textit{C. Metrics:}
Since the ground truth for groupings is not available, we use the standard metrics designed for the quality evaluation of unsupervised clustering approaches with unlabeled data, as follows.\\
\textit{Silhouette Coefficient (SC):} \cite{rousseeuw1987silhouettes} is a value between -1 and 1 that represents the degree of separation between the resulted clusters.
Ideally, the closer the silhouette coefficient is to the number 1, the more high quality of the clustering would be.
The silhouette score is calculated using the mean of the distance inside a cluster and the mean nearest-cluster distance for each sample.\\
\textit{Calinski-Harabasz Index (CH):}\cite{calinski1974dendrite} is a criterion for evaluating clustering techniques that lack ground truth labels (alternatively called the Variance Ratio Criterion, CH score). The CH score indicates how similar an object is to its cluster compared to other clusters. This is accomplished by comparing the distances between objects and their cluster centroids and the distances between cluster centroids and the centroid of all data points. The higher the CH index value, the more pure the clusters, the more distinct the clusters are, and, therefore, the more high quality.

\textit{D. Procedure:}
The overall procedure is illustrated in Figure~\ref{fig:procedure}. As discussed in the data collection section, we considered logs of three time windows at Westermo and used them as input for the vectorization, both with and without pre-processing.
In vectorization, we use the following language models with their specific settings.\\
- tfidf with one to three n-gram range coupled with dimensionality reduction via PCA\\
- The pre-trained FastText models, trained on one million words from Wikipedia on sub-word level\\
- The pre-trained BERT-base uncased model with 12-layer, 768-hidden, 12-heads, and 110M parameters\\
For each of the vector groups, we use K-Mean, Hierarchical Agglomerative, DBSCAN, and Spectral clustering with the best parameters computed via the Elbow method and Greedy search.
After clustering, the specific evaluation metrics are computed and normalized for better reporting.
Key phrases from all the clusters are extracted and then visualized as word clouds.

A focus group study was planned to collect qualitative insights about the results, following the guidelines by Breen et al.~\cite{breen2006practical} and supplemented it with our experience in conducting and reporting focus groups in industry~\cite{abbas2022relationship,abbas1,Abbas5956} (see the bottom half of Figure~\ref{fig:procedure}).
An instrument was drafted and subsequently validated and refined in a pilot session with one expert from the company.
The instrument contained seven questions regarding the perception of the correctness of the results, challenges associated with LogGrouper adoption, and its benefits
A sub-set of the results---containing the clusters and IDs of their elements with key phrases and word cloud---for three time windows was sent to the participants one day before the actual session for review.
The goal of sending the results document to the participants was not to evaluate any specific measure but rather to raise discussions about the general perception of correctness.
The actual session was conducted with eleven experts from the company, including one of the authors of this paper.
The experts were chosen for diversity in roles and experience.
In a short presentation, the goal of the study was introduced, and the LogGrouper API was demoed, followed by the actual session.
The session was recorded after consent was obtained from the experts and then transcribed while anonymizing confidential data~\cite{strandberg2019ethical}.
Following the guidelines proposed by Braun and Clarke~\cite{braun2006using}, thematic analysis was conducted on the transcript and was validated by three authors of this paper.
After the thematic analysis, the audio files were deleted for confidentiality reasons.
The final themes and sub-themes resulted in the qualitative results, reported in Section~\ref{sec:results}.
\begin{figure}[!t]
      \centering
      \includegraphics[width=0.90\linewidth]{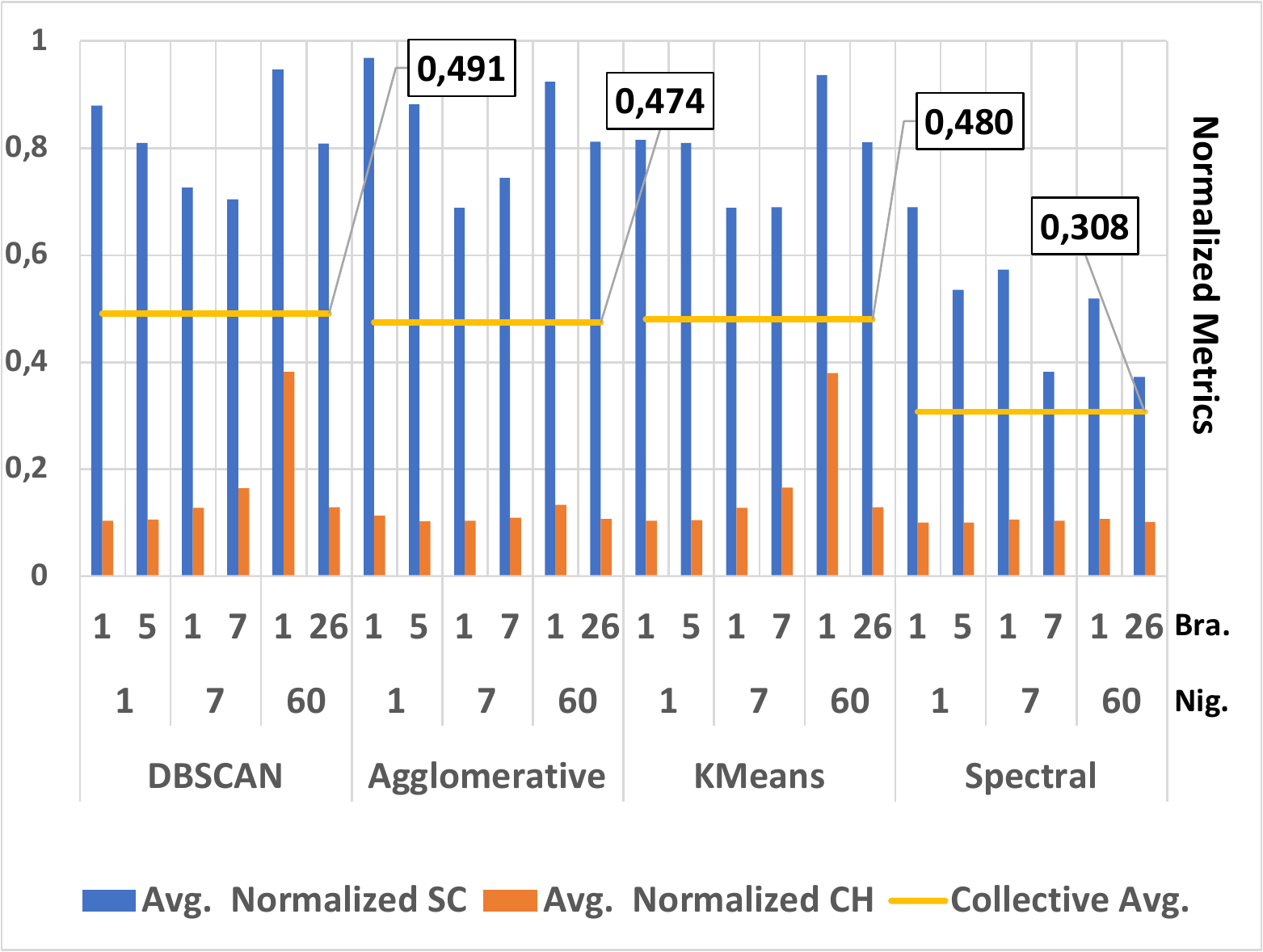}
      \caption{Holistic view of the normalized averaged results}
      \label{fig:results}
 \end{figure}
\section{Results \& Discussion}\label{sec:results}
\subsection{Quantitative Results (RQ1)}
\begin{table*}[]
\centering
\caption{Clustering quality comparison for selected algorithms across one week for one branch}
\label{tab:results}
\begin{tabular}{|c|c|c|c|c|c|}
\hline
\textbf{Clustering}       & \textbf{Vectorizer} & \textbf{SC} & \textbf{CH} & \textbf{ANSC}          & \textbf{ANCH}          \\ \hline
\multirow{6}{*}{DBSCAN}   & BERT                & 0,915       & 2481,469    & \multirow{6}{*}{0,909} & \multirow{6}{*}{0,133} \\ \cline{2-4}
                          & pBERT               & 0,905       & 2296,911    &                        &                        \\ \cline{2-4}
                          & FastText            & 0,798       & 1766,108    &                        &                        \\ \cline{2-4}
                          & pFastText           & 0,811       & 1607,540    &                        &                        \\ \cline{2-4}
                          & TFIDF               & 0,799       & 815,786     &                        &                        \\ \cline{2-4}
                          & pTFIDF              & 0,687       & 543,144     &                        &                        \\ \hline
\multirow{6}{*}{\begin{tabular}[c]{@{}c@{}}Agglome-\\ rative\end{tabular}} & BERT & 0,867 & 42439,533 & \multirow{6}{*}{\textbf{0,939}} & \multirow{6}{*}{\textbf{0,381}} \\ \cline{2-4}
                          & pBERT               & 0,850       & 32037,364   &                        &                        \\ \cline{2-4}
                          & FastText            & 0,848       & 1535,555    &                        &                        \\ \cline{2-4}
                          & pFastText           & 0,811       & 1607,540    &                        &                        \\ \cline{2-4}
                          & TFIDF               & 0,884       & 1503,334    &                        &                        \\ \cline{2-4}
                          & pTFIDF              & 0,867       & 1419,293    &                        &                        \\ \hline
\multirow{6}{*}{kMeans}   & BERT                & 0,857       & 42268,074   & \multirow{6}{*}{0,925} & \multirow{6}{*}{0,380} \\ \cline{2-4}
                          & pBERT               & 0,850       & 32037,364   &                        &                        \\ \cline{2-4}
                          & FastText            & 0,845       & 1500,804    &                        &                        \\ \cline{2-4}
                          & pFastText           & 0,728       & 1402,890    &                        &                        \\ \cline{2-4}
                          & TFIDF               & 0,884       & 1503,334    &                        &                        \\ \cline{2-4}
                          & pTFIDF              & 0,867       & 1419,293    &                        &                        \\ \hline
\multirow{6}{*}{Spectral} & BERT                & -0,133      & 506,165     & \multirow{6}{*}{0,366} & \multirow{6}{*}{0,106} \\ \cline{2-4}
                          & pBERT               & 0,077       & 617,049     &                        &                        \\ \cline{2-4}
                          & FastText            & -0,060      & 31,014      &                        &                        \\ \cline{2-4}
                          & pFastText           & 0,388       & 197,845     &                        &                        \\ \cline{2-4}
                          & TFIDF               & 0,607       & 621,170     &                        &                        \\ \cline{2-4}
                          & pTFIDF              & 0,199       & 66,624      &                        &                        \\ \hline
\end{tabular}
\end{table*}

In this sub-section, we present and discuss the quantitative results. We calculate the evaluation metrics for each of the unique combinations of time windows, vectorization approaches, and clustering algorithms.
However, due to space limitations, we only report the results for a one-week time window for one branch as an example in Table~\ref{tab:results}.
Nevertheless, we provide a holistic view of the collective averaged results across all the combinations in Figure~\ref{fig:results} both for input with and without pre-processing applied.
Furthermore, we provide all the additional results in a spreadsheet for interested readers~\footnote{Replication package at \url{https://doi.org/10.5281/zenodo.6405969}}.
\begin{table}[t!]
\centering
\caption{Themes extracted from the transcript}\label{tab:themes}
\resizebox{\columnwidth}{!}{%
\begin{tabular}{|l|l|}
\hline
\textbf{Theme} &
  \textbf{Sub-theme and codes} \\ \hline
1. Benefits &
  \begin{tabular}[c]{@{}l@{}}1.1 Provides an overview that enables quick reaction\\ to failure assignment\\ 1.2 Highlights diverse failures that otherwise would\\ have been missed\\ 1.3 Could point to cause and enable improved logging\\ 1.4 Aid analysis for detection of intermittent failures\end{tabular} \\ \hline
\begin{tabular}[c]{@{}l@{}}2. Perception of \\correctness\end{tabular} &
  \begin{tabular}[c]{@{}l@{}}2.1 Provides essential grouping that is more important \\than the presentation\\ 2.2 False positives may reduce trust in LogGrouper\end{tabular} \\ \hline
\begin{tabular}[c]{@{}l@{}}3. Challenges \& \\improvements\end{tabular} &
  \begin{tabular}[c]{@{}l@{}} 3.1 Link to other artifacts, such as commits and code\\ 3.2 Historic information of grouping and parameter\\ control for end-users \\3.3 Integration and maintenance of the tool\end{tabular} \\ \hline
\end{tabular}
}
\end{table}

Table~\ref{tab:results} shows the results of the failure log clustering for one week on one code branch.
The ANSC column shows the average normalized score of SC across the vectorization approaches.
In addition, the ANCH column shows the normalized averaged CH score across the vectorization approaches.
Vectorization approaches starting with `p' indicate that pre-processing was applied.
From the results, as we can see in Table~\ref{tab:results}, for one week for one code branch, the Agglomerative clustering approach seems to be performing the best in terms of the normalized averaged SC and CH.
However, the Agglomerative clustering approach is followed very closely by K-Means and DBSCAN.
In different time windows and code branches, we saw that KMeans, DBSCAN, and Agglomerative performed significantly closer to each other in terms of the normalized SC and CH metrics.
In contrast, the Spectral clustering approach performed the worst in the context of our dataset in all of the time windows with a significant difference from other approaches.
Furthermore, on average, both SC and CH metric shows that pre-processing results is clustering with low quality.
Most log messages are typically short and are written in a more formal language following a structure. Therefore, we observed that pre-processing might result in lower quality grouping than when raw log messages are clustered.
This phenomenon of limited vocabulary is also observed in other domains, such as requirements engineering~\cite{8049173}.
Our qualitative evaluation (see the following sub-section) shows that experts prefer pre-processed groups summary over without stop-words.
Nevertheless, in our context, on average, pre-processed input to clustering methods negatively affects the quality of clustering in terms of SC and CH.
We believe that pre-processing can be applied after the actual clustering for presenting summaries as word clouds and key phrases.

We also present a holistic view of the collective average of our evaluation metric over the three time windows in Figure~\ref{fig:results}.
In Figure~\ref{fig:results}, the x-axis shows the number of available code branches (Bra. row in Figure~\ref{fig:results}) in the time windows (Nig. row in Figure~\ref{fig:results} representing the number of nights).
As it can be seen in Figure~\ref{fig:results} that on average (yellow line), the normalized SC and CH is better for the LogGrouper pipeline with the DBSCAN clustering approach.
Furthermore, like our weekly results in Table~\ref{tab:results}, Agglomerative and KMeans clustering approaches follows the DBSCAN approach very closely.
However, the Spectral approach shows the worst results in our particular context.

Based on our quantitative result, we summarize an answer to our RQ1 as follows.
\begin{mdframed}[style=style1]
\textbf{\textit{Summary for RQ1:}} On average, the DBSCAN clustering approach performs the best in grouping failure logs with a combined average of normalized SC and CH score of 0,491.
However, depending on the time window and number of code branches, Agglomerative and KMeans may also outperform DBSCAN in some cases.
In our case, the Spectral clustering approach performed the worst in terms of our evaluation metrics.
Pre-processing appears to have a negative effect on the quality of failure clustering.
\end{mdframed}

\subsection{Qualitative Results (RQ2)}
After the thematic analysis, the extracted themes and sub-themes were reviewed, logically re-ordered, and reported in Table~\ref{tab:themes}.
In this sub-section, we present and discuss the qualitative results with supporting quotes---re-written for better readability---from the transcript presented in \say{italic text} for emphasis.
We also linked the text of this sub-section with the sub-themes presented in Table~\ref{tab:themes} in bold text, as \textbf{(sub-theme number)}.

\subsubsection{Perceived benefits}
The experts in the focus group agreed that the results of LogGrouper approach can provide a high-level overview of failures, can lead to improved logging and provide directions for root cause.

\textbf{(1.1).} It was also agreed that such overview of failures helps in quick reaction to failures and can guide developers into knowing what their code changes are breaking in the overall system.
However, developers see no benefit of such an approach when they are working on a single branch.
\say{As a software developer, yes, it would be more of a general guideline (...) [It] can give a quick pointer to understand what my commit broke...}
On the other hand, the testing team saw these results as an enabler for failure assignments to developers.
The high-level failure grouping is seen as very useful and could be used as a source to identify potential teams that can look into the failures.
Furthermore, this grouping is also perceived useful for identifying similar failures that could be assigned to the same teams.
\say{The grouping makes me skip looking at a lot of log files (...) Is it something in the test framework? or test system? I mean this barrier of which team should take a look at this failure is the first thing (...) sometimes if we have a number of the test failing, you are looking into [assigning] one, and then suddenly someone says can you look into these other similar failures but if I have the clusters, I would assign once.}

\textbf{(1.2).} Typically, nightly failure logs might contain a large number of log messages and experts might skip reviewing a few logs at the end of the list.
However, clustering failure into a limited number of groups aids in highlighting a diverse set of failures and encourages engineers to look into multiple groups.
Experts in the studied setting also agreed that LogGrouper might aid in highlighting the diverse failure groups and, in turn, lead to not missing the last logs at the end.
\say{In the morning, if I have thirty new logs to look at, and 29 of them are the same, maybe I'll look at the first three, and then I skip the rest; then I would miss the different one. [LogGrouper could be] useful because I can see the clusters instead, and I won’t miss the last one.}

\textbf{(1.3).} Furthermore, the groupings are also seen as an initial pointer for common/root cause analysis by the experts.
However, for the root cause analysis, the actual logs are seen as more important than their grouping or the word cloud representing the groups.
\say{From that word cloud, we understood directly what the error was; however, finding the root cause may require looking into the actual log to reproduce the error.}
Experts also highlighted that logs are only as good as they are.
The higher quality the log messages are, the more meaningful the grouping could be.
Therefore, clustering failures in an industrial context might lead to improving logging.
\say{Logs are only as good as they are. If we don't print good logs or good phrases into the logs, it won’t do anything [good for the clustering].
It might also be a good indication that you do not have sufficient logs.
}

\textbf{(1.4).} Finally, experts also highlighted that LogGrouper could be useful in the detection of intermittent failures.
However, for that, LogGrouper must keep historical information on the groupings.
Experts believe that the failure groups over time could be correlated with finding intermittent failures.
\say{We still have problems with intermittent failures. So it's a very good question how these clusters work overtime. Suppose we have one cluster that does not occur very often, maybe not [intermittent]. So, I think that's something that this could be very useful for.}

\subsubsection{Perceived correctness} \textbf{(2.1).} The above-discussed benefits should be realized with correctness at the core.
Therefore, we asked the experts their views on the correctness of the results produced by LogGrouper.
Experts agreed that the initial grouping produced by LogGrouper is more important than the visualization of groups.
Nevertheless, one expert pointed out that the word cloud representation of the group could be a direct initial pointer to the test.
\say{If we look at the phrase here, it's rather an overview of the [feature of a communication protocol], I directly understood which test case it was...}

\textbf{(2.2).} However, experts highlighted that false positives and fuzziness in the grouping could reduce trust in LogGrouper.
In this regard, several improvements were also highlighted (see challenges and improvement theme).
When asked about the preference between fuzzy and strict lexical grouping, experts preferred more strict lexical grouping on failure logs for one night and more nuance semantic groups for a larger time window.
\say{False positives might result in spending more time on something that you might have figured out faster if you weren't sidetracked. (...) I think the biggest adoption problem would be trust...
We need more fuzzy grouping over time but strict grouping for one night.}

\subsubsection{Challenges, limitations and future improvements}
Among many other challenges, here we reported the most discussed challenges and potential improvements suggested by experts for LogGrouper.

\textbf{(3.1).} The log groupings alone are not very specific and could be improved to enable a more detailed root cause analysis by linking other artifacts in the DevOps environment.
For example, experts suggested the extraction of frequent location in the source code that causes most of the failures in a group to be linked to the group.
Experts believe that this will guide the developers and testers in the next steps of root cause analysis and failure reproduction.
\say{[From developers perspective] you would always want to look at the source and if you could link that from here, that would be good.}
Furthermore, an interesting point that the experts suggested was linking the failure groups to commits.
However, this might not be as trivial as test selection process of Westermo~\cite{strandberg2016experience}
might skip running many test cases that could have triggered a failure in a recent commit, and then later, after the commit is merged with the branch, a new test selection might trigger the failure.
That would complicate pinpointing the potential commit causing the failure, as there could be multiple commits between the actual commit causing the failure and the time the failure is triggered.
Nevertheless, a list of a potential commits could still be linked to the failure groups and may provide useful direction for root cause analysis.
\say{We would like to link the causes to commits because that's how we actually do it when we analyze for finding the root causes.}

\textbf{(3.2).} The current version of LogGrouper does not keep any historic grouping information and neglects the timely nature of failure logs.
However, historic grouping information could aid the detection of intermittent failures and may also enable failure and solution pair extraction.
Nevertheless, Westermo already has a solution for the detection of intermittent failures~\cite{strandberg2020intermittently}, but their solution could be complemented with a history of failure grouping.
Furthermore, the distributed nature of the system under study and the timely nature of log files also require additional time-series visualization of the failure logs and their groups.
Experts believe that LogGrouper can be improved by complementing the grouping with grouping history and a time series visualization of the failures.
\say{There is no history in this, but history in LogGrouper would be very useful [for intermittent failures]...  [Historical information can provide a] better understanding of what might actually be a really good group over time.}

\textbf{(3.3).} However, the adoption of LogGrouper is very much dependent on where it is integrated.
The company has many tools for test results visualization and DevOps operations.
The tool where the LogGrouper API will be integrated might have an impact on the potential use in the studied setting.
Furthermore, after the integration, the tool also has to be maintained.
Therefore, a team has to take ownership of the LogGrouper API and maintain the running environment and source code of LogGrouper.
\say{[For the adoption] it depends on how we integrate [LogGrouper]. [Would it be a] separate tool or inside our own tools? I think that will impact how much we will use it. [Regarding integration] we would sort of hold the source code somehow or have access to it, and then some team here would have the ownership (...) we need to assign it to somebody because otherwise, it's just there, but nobody [will use and maintain it.]
}

Based on the qualitative results, we summarized an answer for RQ2, as follows.
\begin{mdframed}[style=style1]
\textbf{\textit{Summary for RQ2:}} Failure log grouping in an industrial DevOps environment could aid failure assignment and act as an initial pointer for root cause analysis. The groupings produced by LogGrouper are perceived as useful. However, historical information, linking failure groups to commits, and tool integration are seen as challenging limitations and must be improved for the adoption of the tool in the industry.
\end{mdframed}

\section{Validity Threats}\label{sec:validity}
In this section, we present the relevant validity threat according to the classification proposed by Runeson and Hoest~\cite{Runeson2009}.

\textit{Construct validity} threats are related to the understanding of phenomena under study by the researchers. 
In this regard, we oriented the quantitative evaluation of LogGrouper on the best-performing clustering algorithm in terms of the quality of clustering.
However, the results are also dependent on the representation vectors that are used as input to the clustering approaches.
Nevertheless, we argue that evaluation of the vectorization approaches together with the clustering algorithm would be very interesting but would require access to ground truth data about grouping and could be interesting for future work.

\textit{Internal validity} threats are related to the credibility of our results.
To mitigate potential internal validity threats, we followed standard open-source implementations of the algorithms and followed standard guidelines for designing the evaluation.
The focus group instrument was also validated in a pilot session.
Furthermore, we included authors with diverse backgrounds and industry experts in the study design.

\textit{External validity} threats relate to the generalizability of our results.
This study is conducted within one Swedish company, developing an operating system for routers and switches.
We believe that one company might not be a good sample. However, we argue that companies with large regression suites producing a huge number of nightly logs could still benefit from the results.
Nevertheless, we do not claim the generalizability of our results beyond the studied context.

\section{Conclusion and Future directions}\label{sec:conc}
This paper presents the LogGrouper approach for failure grouping in an industrial DevOps environment.
Results in our context show that, on average, the density-based clustering approach DBSCAN performed better in terms of the quality of the grouping.
However, KMeans and Agglomerative clustering approaches followed the DBSCAN closely. 
Furthermore, pre-processing, specifically stop word removal, has a negative impact on the quality of the clustering.
Qualitative results suggest that the failure groups produced by LogGrouper are useful and can also be a starting point for root cause analysis and failure assignment.
In addition, failure grouping is seen to highlight diverse groups of failures that might have been skipped by engineers.

However, several challenges and limitations may hinder the adoption of LogGrouper in the industry.
We plan to address some of these challenges and extend the LogGrouper approach for failure and solution pair extraction, among other future directions.
In addition, we plan to explore the use of historic clustering information for the detection of intermittent failures. 

\textit{\textbf{Acknowledgement.}}
This work is partially funded by the AIDOaRt (KDT) and SmartDelta (ITEA) projects.

\bibliographystyle{abbrv}
\begin{small}
\bibliography{references}
\end{small}

\end{document}